\documentclass[12pt,english]{article}
\usepackage[varg]{txfonts}
\usepackage[T1]{fontenc}
\usepackage{textcomp}
\usepackage{amstext}
\usepackage{amssymb}
\usepackage{graphicx}

\makeatletter

\newcommand{\lyxmathsym}[1]{\ifmmode\begingroup\def\b@ld{bold}
  \text{\ifx\math@version\b@ld\bfseries\fi#1}\endgroup\else#1\fi}

\@ifundefined{textcolor}{}
{%
 \definecolor{BLACK}{gray}{0}
 \definecolor{WHITE}{gray}{1}
 \definecolor{RED}{rgb}{1,0,0}
 \definecolor{GREEN}{rgb}{0,1,0}
 \definecolor{BLUE}{rgb}{0,0,1}
 \definecolor{CYAN}{cmyk}{1,0,0,0}
 \definecolor{MAGENTA}{cmyk}{0,1,0,0}
 \definecolor{YELLOW}{cmyk}{0,0,1,0}
}

\makeatother

\usepackage{babel}
\begin{document}

\title{Quantum computing and the brain: quantum nets, dessins d'enfants and neural networks}

%\author{
%\firstname{Torsten} \lastname{Asselmeyer-Maluga}\inst{1}\fnsep\thanks{\email{torsten.asselmeyer-maluga@dlr.de}} 
%}
\author{Torsten Asselmeyer-Maluga \\ German Aerospace Center, Rosa-Luxemburg-Str. 2, 10178 Berlin, Germany \\ torsten.asselmeyer-maluga@dlr.de}
%\institute{German Aerospace Center, Rosa-Luxemburg-Str. 2, 10178 Berlin, Germany}
\maketitle

\abstract{%
In this paper, we will discuss a formal link between neural networks
and quantum computing. For that purpose we will present a simple model
for the description of the neural network by forming sub-graphs of
the whole network with the same or a similar state. We will describe
the interaction between these areas by closed loops, the feedback
loops. The change of the graph is given by the deformations of the
loops. This fact can be mathematically formalized by the fundamental
group of the graph. Furthermore the neuron has two basic states $|0\rangle$
(ground state) and $|1\rangle$ (excited state). The whole state of
an area of neurons is the linear combination of the two basic state
with complex coefficients representing the signals (with 3 Parameters:
amplitude, frequency and phase) along the neurons. If something changed
in this area, we need a transformation which will preserve this general
form of a state (mathematically, this transformation must be an element
of the group $SL(2;\mathbb{C})$). The same argumentation must be
true for the feedback loops, i.e. a general transformation of states
along the feedback loops is an assignment of this loop to an element
of the transformation group. Then it can be shown that the set of
all signals forms a manifold (character variety) and all properties
of the network must be encoded in this manifold. In the paper, we
will discuss how to interpret learning and intuition in this model.
Using the Morgan-Shalen compactification, the limit for signals with
large amplitude can be analyzed by using quasi-Fuchsian groups as
represented by dessins d'enfants (graphs to analyze Riemannian surfaces).
As shown by Planat and collaborators, these dessins d'enfants are
a direct bridge to (topological) quantum computing with permutation
groups. The normalization of the signal reduces to the group $SU(2)$
and the whole model to a quantum network. Then we have a direct connection
to quantum circuits. This network can be transformed into operations
on tensor networks. Formally we will obtain a link between machine
learning and Quantum computing.
}
%\maketitle

\section{Introduction}

Our brain (which will be seen as a neural network in the following)
is known to be a complex system with no chance to describe it at least
partly. The reason for this is simple The complexity of a neural network
is mainly given by the exponential size of the state space. Interestingly,
a quantum circuit has also an exponential state space as usually given
by $2^{N}$ where $N$ is the number of qubits. This paper will be
discuss the question whether both systems have a similar description.
This idea sounds crazy at the first view. But the brain will be viewed
as a classical computer with an exponential state space. In the brain,
the so-called calculations are given by the signals going through
the network. The number of possible ways in a network is exponential
large with respect to the size (number of vertices) of the network.
A classical computer can be simulate a quantum computer but needs
exponential ressources and a far longer calculation time. The motivation
for this work is similar idea: is it possible that the brains (or
a neural network) can be simulate a quantum computer with smaller
ressources?

Here we will discuss this idea. But in the following we don't discuss
a direct link between neural networks and quantum computing. Therefore
we don't assume things like entangled neurons or similar things. 
Furthermore we also don't discuss the interaction between machine learning 
and quantum computing (see \cite{BriegelDunjkoTaylor2016,Briegel2017}). But
the main problem of this work is the question: how can we describe
such a complex system like the brain?

In physics there were two substantial revolutions: relativity and
quantum theory changing our view of nature. In principle, there was
also a third revolution which can be called 'Understanding complexity'
with topics like deterministic chaos, self-organization, synergy etc.
Partly implied by this progress, other branches of natural science
like biology and chemistry were developed also much further in particular
at the second half of this century. In parallel, mathematics changed
also their shape, away from analysis and analytical geometry more
in the direction of qualitative methods. Here, the development of
topology was central for this direction influencing many areas like
algebraic geometry, number and operator theory. Topology can be simply
expressed as a qualitative theory of shapes and spaces. So, why not
using these ideas to grasp complex systems? This paper used these
methods to fill this gap. For that purpose we will concentrate on
one of the most complex system, the human brain. Currently there is
a big effort to understand it in the Human Brain Project (a EU FET
Flagship about 1 billion Euro expensive). But is it possible to understanding
processes like learning or intuition without a large simulation of
the whole brain? Maybe a combination of advanced physics and mathematics
can help. A basic model of neural networks is the famous Ising model.
But in the last two or three decades much progress was made to describe
spacetime using graphs (the so-called spin nets of Penrose). These
graphs are described by quantum mechanics (and are part of quantum
gravity). Therefore there is 'room' for changes in this graph by quantum
fluctuations. This picture has much in common with the human brain
with fluctuations in the neural network (leading to connection changes
in the network). So why not use these spin nets to describe the neural
network? In this paper we will discuss this possibility by describing
a simple model. One guiding principle of this approach is the interaction
between connected neurons to obtain a non-trivial model. Usually not
only a pair of neurons is involved into an interaction but rather
a subset of neurons. We will describe this interaction as feedback
loops. Below we will show that this model has features which one would
expect (or know) from neural networks. The mathematics behind this
approach based on the work of Morgan and Shalen \cite{MorganShalenI1984}
which originally described hyperbolic 3-manifolds (known as Morgan-Shalen
moduli space compactification).

In the next section, we will describe the basic model. The brain can
be seen as a dynamical graph with electrical signals having amplitude,
frequency and phase. Because of the complexity of the graph, it is
hopeless to include the whole graph. Instead we form areas of neurons
having the same state (ground state or excited state). We describe
the interaction between these areas by closed loops, the feedback
loops. The change of the graph is given by deformations of the loops.
At first view, the interaction between the neurons in the area as
represented by loops cannot be neglected. It is equivalent to say
that the loops cannot be contracted by deformations. This fact can
be mathematically formalized by the fundamental group of the graph.
Furthermore the neuron has two basic states $|0\rangle$ (ground state)
and $|1\rangle$ (excited state). The whole state of an area of neurons
can be obtained fomally by suming over all ground and excited states
including the corresonding amplitude. Then it is the linear combination
of the two basic state with complex coefficients representing the
signals (with 3 Parameters: amplitude, frequency and phase) along
the neurons. If something changed in this area, we need a transformation
which will preserve this general form of a state (mathematically,
this transformation must be an element of the group $SL(2;\mathbb{C})$).
The same argumentation must be true for the feedback loops, i.e. a
general transformation of states along the feedback loops is an assignment
of this loop to an element of the transformation group. Then it can
be shown that the set of all signals forms a manifold (character variety)
and all properties of the Network must be encoded in this manifold.
In the section \ref{sec:The-dynamics-of}, we will discuss how to
interpret learning and intuition in this model by using the Morgan-Shalen
compactification. In particular, the limit for large amplitudes of
the signals can be analyzed by using quasi-Fuchsian groups as represented
by dessins d'enfants (graphs to analyze Riemannian surfaces). As shown
by Planat and collaborators, these dessins d'enfants are a direct
bridge to (topological) quantum computing with permutation groups.
The structure of the quasi-Fuchsian groups gives a measure of the
complexity as shown in section \ref{sec:On-the-complexity}. In the
section \ref{sec:The-relation-to}, we will discuss the normalization
of the signal seen as reduction to the group $SU(2)$. It reduces
the whole model to a quantum network but only for a fixed decomposition
of the network into areas. Then we have a direct connection to quantum
circuits. Then we can define operations on tensor networks. Formally
we will obtain a link between machine learning and Quantum computing.

\section{A simple model of a neural network\label{sec:A-simple-neural} }

Neural networks as used for deep learning / machine learning are based
on the Ising model. Our model is inspired by this model but we will
implicitly also use the spin networks known from quantum gravity.
It is a difficult task to create a model which is complex enough to
generate a realistic behavior but simple enough to work with. For
the following, let us assume a graph representing the network of neurons.
This graph is non-planar, i.e. the graphs needs a 3-dimensional space
to embed them without any intersections of edges. At first we will
fix this graph (or network) but later we will allow for fluctuations
of the graph (i.e. the change of the edges). Here we are mainly interested
in the flow through this network. In the following we will use the
words network and graph as synonyms. It is known that:\\
\emph{There is substantial evidence that a \textquotedbl{}top-down\textquotedbl{}
flow of neural activity (i.e., activity propagating from the frontal
cortex to sensory areas) is more predictive of conscious awareness
than a \textquotedbl{}bottom-up\textquotedbl{} flow of activity.}\\
This fact implies a hierarchical structure inside the network. Then
there are areas (part of the network) with a similar activity which
are separated from other areas. The whole structure of the network
is controlled by another area where this area is connected to many
other areas to control them. There are many ways to send a signal
from one point in the network to another point. Usually all these
ways are used by the signals. The network is very large so that the
corresponding area can be seen as a continuous space in good approximation.
The whole approach can be seen as a kind of cluster approximation:
the network is clustered into areas. Let $\mathcal{N}$ be this network
which is decomposed into areas (or sub-networks) $G_{i}$with
\[
\mathcal{N}=\bigcup_{i}G_{i}
\]
Main idea in the approach is the observation that there is an interaction
between neurons, i.e. a signal from one neuron will return directly
or after a greater loop. These signals are signals forming a loop
which we call \emph{feedback loops}. This loop can be seen as a real
interaction between some neurons (or areas of neurons). Usually in
an area, there is more than one feedback loop. But the network is
not fixed and now we will discuss fluctuations of the network. Usually
neurons build new connections and/or remove some other. Because of
this behavior some of the feedback loops will be destroyed or new
loops will be created. In contrast, there are stable feedback loops
which are not destroyed (because they are used very often). This behavior
is comparable to contractable and non-contractable loops in topology.
Consider a loop in a disk which can be arbitrarily deformed and finally
contracted to a point. In contrast, a disk with a hole contains contractable
loops (= all loops not going around the loop) and non-contractable
loops (= all loops going around the hole). All non-contractable loops
are characterized by the number how often the loop winds around the
hole. But more importantly, the loop going around the hole is stable
with respect to any deformation (or fluctuation) which don't destroy
the hole in the disk (or producing a new hole). Therefore we will state: the feedback loops
produce the topology of the network. 

Closed loops are forming a monoid with concatenation as operation.
This monoid can be completed to a group by using the deformation of
loops by using the concept of homotopy (one-parameter family of deformations).
This group is known as fundamental group $\pi_{1}(\mathcal{N})$ of
$\mathcal{N}$, see \cite{GrHa:81,Nak:89} for details. Usually this
group is generated by a finite number of generators $w_{1},\ldots,w_{n}$
forming sequences (the group operation) which are restricted by relations
$r_{1},\ldots,r_{m}$. Then $\pi_{1}(\mathcal{N})$ is given by
\[
\pi_{1}(\mathcal{N})=\langle w_{1},\ldots,w_{n}|\, r_{1},\ldots,r_{m}\rangle
\]
and we will discuss some examples now. The free group $\langle a|\emptyset\rangle$
of one generator and no relation are sequences $a^{k}$ for every
$k\in\mathbb{Z}$, i.e. $\langle a|\emptyset\rangle$ is isomorphic
to $\mathbb{Z}$. In a similar manner, the free group $\langle a,b|\emptyset\rangle$
is isomorphic $\mathbb{Z}\star\mathbb{Z}$ to a free product. We will
later see that this group can be represented by a tree (the Cayley
graph of this group). The group $\langle a|\, a^{q}\rangle$ with
one generator and one relation $a^{q}=e$ ($e$ unit of the group)
are sequences $a^{k}$ with $-q<k<q$, i.e. this group is isomorphic
to $\mathbb{Z}_{q}$. Other examples are $\langle a,b|\, aba^{-1}b^{-1}\rangle=\mathbb{Z}\oplus\mathbb{Z}$
with relation $aba^{-1}b^{-1}=e$ or $ab=ba$ which is the fundamental
group $\pi_{1}(T^{2})$ of the torus.

Now we will reverse the argumentation, i.e. for a given group $G$
with finitely many generators we have to consider a space/manifold $X$ so
that $G=\pi_{1}(X)$. Using the cellular approximation theorem, the
space $X$ must be a 2-complex. This 2-complex consists of 1-cells
and 2-cells. The 1-cells give the generators of $G$ and the 2-cells
represent the relations in $G$. It is a result of Curtis \cite{Curtis1962}
that any 2-complex is homotopy equivalent to a 2-complex which embeds
in $\mathbb{R}^{4}$. This fact is the reason that every finitely
generated group can be realized as the fundamental group of a 4-manifold. 

The fundamental group $\pi_{1}(\mathcal{N})$ represents the interaction
between the neurons and $\mathcal{N}$ denotes now the interaction
graph of the neural network. This graph is similar to the neural network,
i.e. this graph can be embedded into a 3-dimensional space. Following
this argumentation, the 2-complex must be also embedded into a 3-dimensional
space or $\pi_{1}(\mathcal{N})$ is the fundamental group of a 3-manifold
also denoted as $\mathcal{N}$. By now it is only a conjecture but
for the following line of arguments it is unimportant i.e. abstractly
we have a 3- or 4-manifold denoted by $\mathcal{N}$ with fundamental
group $\pi_{1}(\mathcal{N})$. The choice of this manifold us unique
by deep theorems (solution of the Poincare conjecture for topological
3- and 4-manifold by Perelman \cite{Per:02,Per:03.1,Per:03.2} and
Freedman \cite{Fre:82}). 

Furthermore we have a function over this space $\mathcal{N}$ which
assigns to every point a state: $|0\rangle$ (ground state) or $|1\rangle$
(excited state). This function is given by the neuron behavior: $|0\rangle$
are signals in $\mathcal{N}$ below the activation of neurons whereas
$|1\rangle$ are signals above the threshold. Every signal is given
by three parameters: strength (or amplitude), frequency and (relative)
phase. These three parameters can be encoded into a complex number
$a\in\mathbb{C}$ with state $a|0\rangle$ or $a|1\rangle$. For an
area $G_{i}$, the state $|\phi(G_{i})\rangle$ is the sum of all
$a|0\rangle$ and $b|1\rangle$. To express this fact mathematically,
we will assign to every area of neurons a state
\[
|\phi\rangle=a\cdot|0\rangle+b\cdot|1\rangle\qquad a,b\in\mathbb{C}
\]
where $a,b$ are complex numbers with a norm $|a|$ representing the
strength of the signal for the state $|0\rangle$ in this area and
$|b|$ the strength for $|1\rangle$. The phase and the signal frequency
of $a$ and $b$ is the relative phase as well the frequency of a
signal going through this area. Here, the usage of complex numbers
is very important, otherwise we are not able to describe the amplitude,
frequency and phase in one structure. The parameter will generate
a 2-dimensional complex vector space $\mathbb{C}^{2}$. A parameter
change is given by the automorphism of this space as given by the
group $SL(2,\mathbb{C})$. If something changed in this area, then
we need a transformation which will preserve this general form of
a state (mathematically, this transformation must be an element of
the group $SL(2,\mathbb{C})$). The same argumentation must be true
for the feedback loops, i.e. a general transformation of states along
the feedback loops is an assignment of this loop to an element of
the transformation group (i.e. a homomorphism $\pi_{1}(\mathcal{N})\to SL(2,\mathbb{C})$
a representation). Some transformations are ruled out, i.e. if we
transform all loops by the same element then we will change nothing
(it is a simple gauge of the network).

These ideas can be expressed mathematically which will be done in
the following. Main part in our argumentation is the inclusion of
neuron interaction by feedback loops. A loop is formally given by
loops of the state $|0\rangle$ and of the state $|1\rangle$. A change
of the loop by a change of the network or by a change of the currents
in the network are given by a map
\[
\pi_{1}(\mathcal{N})\to SL(2,\mathbb{C})
\]
which must be preserve the group structure (i.e. it is a homomorphism).
Of course, all parameters can be gauged at the same time without any
change. This fact can be expressed by an action of the group $SL(2,\mathbb{C})$
on the homomorphism above. This action is given by conjugation (the
action $(g,a)\mapsto g\cdot a\cdot g^{-1}$). Mathematically expressed,
the set of homomorphism will be denoted by 
\[
Hom(\pi_{1}(\mathcal{N}),SL(2,\mathbb{C}))=\left\{ \pi_{1}(\mathcal{N})\to SL(2,\mathbb{C})\right\} 
\]
which is unique up to conjugation (the action $(g,a)\mapsto g\cdot a\cdot g^{-1}$)
\[
\mathcal{M}=Hom(\pi_{1}(\mathcal{N}),SL(2,\mathbb{C}))/SL(2,\mathbb{C})
\]
and finally we will obtain the space $\mathcal{M}$. The transformation
group $SL(2,\mathbb{C})$ of our parameters (seen as automorphism
of the parameter space as 2D complex vector space) is well-known in
mathematics: it is the isometry group of the 3-dimensional hyperbolic
space (honestly it is the universal cover of this isometry group but
this difference is unimportant here). Then the space $\mathcal{M}$
is the moduli space of hyperbolic structures on $\mathcal{N}$.

Now let us summarize the model:
\begin{enumerate}
\item There is a network of neurons with a hierarchical organization. All
signals going through this network
\item There is a set of feedback loops which are stable with respect to
a change of the network. These stable loops are identified with non-contractable
loops in a 3-dimensional space $\mathcal{N}$. This set of loops has
the structure of a group $\pi_{1}(\mathcal{N})$.
\item Every neuron has a state out of two possibilities: $|0\rangle$ (ground
state) or $|1\rangle$ (excited state). For an area of neurons we
assign a state $|\phi\rangle=a\cdot|0\rangle+b\cdot|1\rangle$ with
the complex numbers $a,b$. The norm $|a|$ represents the strength
of the signal for the state $|0\rangle$ in this area and $|b|$ the
strength for $|1\rangle$. The phase and the signal frequency of $a$
and $b$ is the relative phase as well the frequency of a signal going
through this area.
\item A transformation of the state $|\phi\rangle$ must be end up with
a state again, i.e. the possible transformations forming a group $SL(2,\mathbb{C})$.
Then a transformation of feedback loops is a map which assigns every
loop (an element of $\pi_{1}(\mathcal{N})$) to a transformation (an
element of $SL(2,\mathbb{C})$) respecting the group structures of
the transformation and loops. Gauging of the network is ruled out.
\end{enumerate}

\section{The dynamics of this model: learning and intuition\label{sec:The-dynamics-of}}

The model above is only the frame for a general dynamics. But at first
we will ask what we expect if a sensor like the eye gets a signal which
will be recognized to be a dog by the neural network? In the previous
section we analyzed the structure of the signals inside the network.
Here, feedback loops as expression for neuron interactions are the
most important ingredients for us. Now let us consider two signals
for different objects. One would expect that the endpoints of the
signals (after going through the network) are separated from each
other. Is this behavior realized in our model? If the network has
learned to recognize different objects (where the input signal came
from the eye) then the corresponding ways of a signal should have a maximized
amplitude and frequency for the signal. Therefore our analysis should
be concentrate on the maximized signals inside the networks. 

This idea can be realized mathematically by considering a signal function
\[
F:\mathcal{M}\to\mathbb{C}
\]
with amplitude, frequency and phase over the space $\mathcal{M}$
of all signals as described above. As shown by \cite{MorganShalenI1984},
the space $\mathcal{M}$ can be described by coordinates given by
group characters. Let $\rho:\pi_{1}(\mathcal{N})\to SL(2,\mathbb{C})$
be one representation. The character is defined by $\chi_{\rho}(\gamma)=Tr(\rho(\gamma))$
for a $\gamma\in\pi_{1}(\mathcal{N})$. The set of all characters
forms an algebraic variety which is equivalent to $\mathcal{M}$.
This approach showed that every function over $\mathcal{M}$ must
be given by a character or the signal function itself is a coordinate
function of $\mathcal{M}$. Of course, this function can be also defined
for the areas $F|_{G_{i}}:G_{i}\to\mathbb{C}$. In principle, there
are two kind of signals: signals with a convergent amplitude and signals
with a divergent amplitude. In our model, both signals have different
interpretations. Convergent signals are given for ways which are rarely
used (or which do not activate the neurons). In contrast, divergent
signals are frequently used ways. As far as we know, the divergent
signals are the most important ones because these ways represent the
learned signals way. As motivated above, we expect that these ways
are separated from other signals i.e. ending at different ends. Expressed
differently, the ways of the signal after a learning are forming a
tree where the endpoints represent the different recognized objects.
This tree must be formed by the signals with divergent amplitudes.
Mathematically, we have to understand the compactification of the
space $\mathcal{M}$.

Morgan and Shalen \cite{MorganShalenI1984} studied a compactification
of this space or better they determined the structure of the divergent
signals. The compactification $\overline{\mathcal{M}}$ is defined
as follows: let $C$ be the set of conjugacy classes of $\Gamma=\pi_{1}(\mathcal{N})$,
and let $\mathbb{P}(C)=\mathbb{P}(\mathbb{R}^{C})$ be the (real)
projective space of non-zero, positive functions on $C$. Define the
map $\vartheta:\mathcal{M}\to\mathbb{P}(C)$ by 
\[
\vartheta(\rho)=\left\{ log(|\chi_{\rho}(\gamma)|+2)\:|\,\gamma\in C\right\} 
\]
and let $\mathcal{M}^{+}$ denote the one point compactification of
$\mathcal{M}$ with the inclusion map $\iota:\mathcal{M}\to\mathcal{M}^{+}$.
Finally, $\overline{\mathcal{M}}$ is defined to be the closure of
the embedded image of $\mathcal{M}$ in $\mathcal{M}\times\mathbb{P}(C)$
by the map $\iota\times\vartheta$. It is proved in \cite{MorganShalenI1984}
that $\mathcal{M}$ is compact and that the boundary points consist
of projective length functions on $\Gamma$ (see below for the definition).
Note that in its definition, $\vartheta(\rho)$ could be replaced
by the function $\left\{ \ell_{\rho}(\gamma)\right\} _{\gamma\in C}$,
where $\ell_{\rho}$ denotes the translation length for the action
of $\rho(\gamma)$ on $\mathbb{H}^{3}$ (3D hyperbolic space) 
\[
\ell_{\rho}(\gamma)=inf\left\{ dist_{\mathbb{H}^{3}}(x,\rho(\gamma)x)\,|\, x\in\mathbb{H}^{3}\right\} 
\]
where $dist_{\mathbb{H}^{3}}$ denotes the (standard) distance in
the 3D hyperbolic space $\mathbb{H}^{3}$. 

Recall that an $\mathbb{R}$-tree is a metric space $(T,d{}_{T})$
such that any two points $x,y\in T$ are connected by a segment $[x,y]$,
i.e. a rectifiable arc isometric to a compact (possibly degenerate)
interval in $\mathbb{R}$ whose length realizes $d_{T}(x,y)$, and
that $[x,y]$ is the unique embedded path from $x$ to $y$. We say
that $x\in T$ is an edge point (resp. vertex ) if $T\setminus\left\{ x\right\}$ has
two (resp. more than two) components. A $\Gamma$-tree is an $\mathbb{R}$-tree
with an action of $\Gamma$ by isometries, and it is called minimal
if there is no proper $\Gamma$-invariant subtree. We say that $\Gamma$
fixes an end of $T$ (or more simply, that $T$ has a fixed end) if
there is a ray $R\subset T$ such that for every $\gamma\in\Gamma$
, $\gamma(R)\cap R$ is a subray. Given an $\mathbb{R}$-tree $(T,d_{T})$,
the associated length function $\ell_{T}:\Gamma\to\mathbb{R}^{+}$
is defined by
\[
\ell_{T}(\gamma)=inf_{x\in T}d_{T}(x,\gamma x)
\]
If $\ell_{T}\not=0$, which is equivalent to $\Gamma$ having no fixed
point in $T$ (cf. \cite{MorganShalenI1984,MorganShalenII1988}, Prop.
II.2.15), then the class of $\ell_{T}$ in $\mathbb{P}(C)$ is called
a projective length function.

Now we are able to formulate the main result:

\emph{If $\rho_{k}\in\mathcal{M}$ is an unbounded sequence, then
there exist constants $\lambda_{k}\to\infty$ (renormalization of
the sequence) so that the rescaled length 
\[
\frac{1}{\lambda_{k}}\ell_{\rho_{k}}
\]
converge to $\ell_{\rho_{\infty}}$for $\rho_{\infty}:\Gamma\to Isom(T)$
a representation of $\Gamma$ in the isometry group of the $\mathbb{R}-$tree
$T$, i.e. we have the convergence
\[
\frac{1}{\lambda_{k}}\ell_{\rho_{k}}\Longrightarrow\ell_{T}
\]
}

\begin{figure}
\includegraphics[scale=0.3]{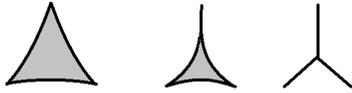}

\caption{degeneration of a triangle to a tree \label{fig:degeneration-of-triangle}}
\end{figure}

But what is the meaning of this result? If the signals in the network
are bounded nothing happens. The signals can follow all possible ways
and there is no separation of signals. This behavior changes for unbounded
signals. According to the result above, the underlying graph for signals
degenerates to a tree. But this tree will give the right behavior,
it separates the signals. This amazing result can be understood on
geometric grounds. The main idea is the usage of the fact that the
assignment of a state to the loop by a map $\pi_{1}(\mathcal{N})\to SL(2,\mathbb{C})$
is a representative of a hyperbolic structure on the 3-manifold $\mathcal{N}$
which is unique up to gauge (the homomorphism $\pi_{1}(\mathcal{N})\to SL(2,\mathbb{C}$
) is determined up to conjugation). The set of all signals is given
by the space $\mathcal{M}$ with a corresponding function $\mathcal{M}\to\mathbb{C}$
for the amplitude, frequency and phase. If the value of this function
restricted to an area grows up over the time then it is likely that
this signal way was learned by the network and it will be stable over
time. Interestingly, one has a direct way to visualize this growing:
the curvature of the hyperbolic space grows up too. But what is effect
on the network. At the vertices, one has the neurons which are connected
by the lines. As one can easily see the lines are concavely curved
(because of the negative curvature). Then the increasing curvature
is given by an increasing curving of the lines. At the limit, the
triangle degenerates and become a tree (see Fig. \ref{fig:degeneration-of-triangle}
), see \cite{MorganShalenIII1988}.
 So, for the maximized signal amplitude (and frequency), the underlying
graph of the signals degenerates to a tree where the endpoints of
the signal are endpoints of the tree (or the leaves of the tree).
Here we will point out that the network itself don't degenerates but
rather the signals on the network. The signal for low amplitudes and
frequencies see the whole graph where the signals for high amplitude
and frequency will see a tree. At the end, the network has learned
to distinguished between objects. In particular the hyperbolicity
of the geometry implies that the distinction is not continuously related
to the start point. Consider for example, two objects say 'table'
and 'chair' which are not so far away from the recognition of the
pattern(three or four legs, some plate, similar size or material etc.).
But we learned to distinguished these two objects.\\
\emph{Learning: the separation of the endpoints for the signals in
the network by the amplitude of the signal}\\
Above we described a recognition of an object as learned by the network.
But how about the process of the learning itself? Here the work of
Morgan and Shalen can help to find an inspiration or better an interpretation
of learning ion our model again. In the approach above, the number
of loops is very important. Usually one needs more then two generators
and some relations between the generators to get this result. The
relations between the loops can be understood as kind of interaction
between the loops (or as global relation). Inspired by this fact,
we will understood learning as a two-step dynamics:
\begin{enumerate}
\item Increase the number of feedback loops together with relations between
them.
\item Increase the amplitudes/frequencies, i.e. repeat the lesson again
and again.
\end{enumerate}
Then in the limit, the amplitudes will increase and one will get the
separation, i.e. the graph will change to a tree. Abstractly spoken,
the learning set is classified into discrete classes where every class
corresponds to the endpoint of the tree. 

For the process 'learning', the number of loops have to be increased.
What happens if we will keep the number of loops fixed but change
the relations between the loops? According to the theory of Morgan
and Shalen, the tree itself will deform and one will get new endpoints,
i.e. the classification of the learning set (used to generate the
tree) changes or the network evaluates the input in a new manner.
But this process is known as intuition (or creation of an idea)\\
\emph{Intuition: generating new relations between the loops by fixing
the loops itself.}\\
Here we will stop for the moment. Our model seems to be a qualitative
model at the first view. But the mathematical model behind this approach
will also allow for a more quantitative approach. But the main problem
is only to fit any realistic data to the parameters of the model (which
will imply a lot of work).

\section{On the complexity of network dynamics and dessin d'enfants\label{sec:On-the-complexity}}

\begin{figure}
\includegraphics[scale=0.2]{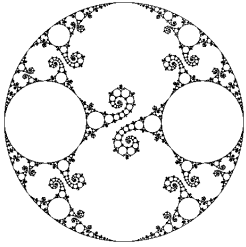} 

\caption{one example of limit (fractal) curves of quasi-Fuchsian groups\label{fig:examples-of-fractal-q-Fuchsian-1}}
\end{figure}

Now we will concentrate on another aspect of this approach: does all
endpoints of the tree forming a kind of pattern from which we can
extract whether the network has already learned the information? Up
to now we don't spoke about the boundary of $\mathcal{N}$, i.e. the
area where the signals start and/or end. The boundary allow also for
loops $\pi_{1}(\partial\mathcal{N})$ which are embedded into $\mathcal{N}$
with transformations $\pi_{1}(\partial\mathcal{N})\to SL(2,\mathbb{C})$.
In principle, we are interested in the signals which approach this
boundary, i.e. we rather consider the 3-manifold $M=\partial\mathcal{N}\times(0,1)$
together with a representation $\rho_{S}:\pi_{1}(\partial\mathcal{N})\to SL(2,\mathbb{C})$.
By using this representation, one can define $M$ as factor $\mathbb{H}^{3}/\rho_{S}(\pi_{1}(\partial\mathcal{N}))$
with metric $dt^{2}+cosh(t)^{2}g$ with the metric $g$ on $\partial\mathcal{N}$.
The manifold $M$ is embedded in $\mathbb{H}^{3}$. In the limit,
the process (seen as orbit of a point under the action of $\rho_{S}(\pi_{1}(\partial\mathcal{N}))$)
(called quasi-Fuchsian group) will approach the boundary at infinity
of $\mathbb{H}^{3}$ given by $\partial_{\infty}\mathbb{H}^{3}=S^{2}$.
Then the limit set is the equator of this $S^{2}$ (a Jordan curve
which divides the sphere into a northern and southern hemisphere),
see \cite{quasi-Fuchsian}. Now there are two possible classes: the
process converges regularly, i.e. the limit set is a smooth circle
or the process converges to a closed fractal curve. In Fig. \ref{fig:examples-of-fractal-q-Fuchsian-1}
we visualizes one examples of this curve for different groups.
 The appearance of a fractal is a sign of complexity. Indeed, the
complexity of the representation $\rho_{S}$ is a direct expression.
But this complexity is related to the complexity of the group $\pi_{1}(\mathcal{N})$,
i.e. the complexity is given by the interaction in the network. In
the other case where the amplitudes are forming a graph (therefore
unable to classify information) the endpoints of the graph are not
forming a fractal curve. So, the fractal curve is a sign for the completed
learning process. 

\begin{figure}
\includegraphics[scale=0.3]{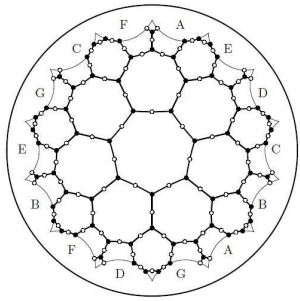}

\caption{dessin d'enfant for a surface (repainted from \cite{dessin-enfant-1})
\label{fig:dessin-d-enfant-surface}}

\end{figure}

The surface at the boundary $\partial\mathcal{N}$ is given by a tree,
see the result of the previous section. It is known that trees are
given by polynomials which are locally given at the branching points
by $z^{n}$ with $n$ branches. These branching points define locally
a ramified covering of the surface over $S^{2}$ (the boundary of
$\mathbb{H}^{3}$ at infinity). These data are enough to define a
dessin d'enfant \cite{dessin-enfant-1}. Interestingly, the corresponding
dessin d'enfant is given by the triangulation of the surface, see
Fig. \ref{fig:dessin-d-enfant-surface} for an example. 
But there is also a second line of arguments leading to the same
result. Let $G\lyxmathsym{\textasciiacute}$ be a subgroup of the
free group $G=\langle a,b|\emptyset\rangle$ endowed with a set of
relations and $H$ a subgroup of $G$ of index $n$. As shown in \cite{Planat2017,Planat2017a},
the permutation representation $P$ associated to the pair $(G\lyxmathsym{\textasciiacute},H)$
is a dessin d\textquoteright{}enfant whose edges are encoded by the
representative of cosets of $H$ in $G\lyxmathsym{\textasciiacute}$.
The Cayley graph of the free group $G$ is a tree and we have again
the relation between the tree at the boundary and dessin d'enfants.

\section{The relation to quantum operations\label{sec:The-relation-to}}

Even the analysis of the previous section uncovers a formal analogy
between our neural network model and quantum computing. At first we
will start with the main assumption: Let $\mathcal{N}$ be the network
with a fixed partitioning into areas $G_{i}$ so that $\mathcal{N}=\bigcup_{i}G_{i}$.
For the description of the state, there are two possibilities: one
can sum up all states of the areas to get one state for $\mathcal{N}$
but at the same time one will loose many important informations or
one forms a tensor state
\[
|\Phi_{\mathcal{N}}\rangle=|\phi_{G_{1}}\rangle\otimes|\phi_{G_{1}}\rangle\otimes\cdots\otimes|\phi_{G_{i}}\rangle\otimes\cdots
\]
to preserve this information. Furthermore this state reflects also
the exponential size of neural network states in general. 

Now it is possible to define operations on this state. For instance,
if the signal amplitudes increases so that all neurons in the area
will change from the ground state $|0\rangle$ to the excited state
$|1\rangle$ then this change can be seen as NOT gate (in the quantum
computing) 
\[
NOT=\left(\begin{array}{cc}
0 & 1\\
1 & 0
\end{array}\right)
\]
acting on the the state $|\phi_{G}\rangle$ of the area. In principle,
one can define every 1-quibit gate in the quantum computing. Of course,
the corresponding state is not a quantum state because the state is
not normalized. But it is an easy task to normalize the state $|\phi_{G}\rangle=a|0\rangle+b|1\rangle$
for every area by dividing the coefficients by the expression $\sqrt{|a|^{2}+|b|^{2}}$
. Then every element of $SU(2)$ will preserve this state. A similar
argumentation can be used to introduce 2-qubit operations. Here one
needs an interaction (or coupling) between the two areas, say $G_{1}$
and $G_{2}$, with a common state $|\phi_{G_{1}}\rangle\otimes|\phi_{G_{2}}\rangle$.
Now one can choose the coupling so that the excited state $|1\rangle$
of $|\phi_{G_{1}}\rangle$ activates the neurons in $G_{2}$ which
is nothing as the action of the $NOT$ gate on $|\phi_{G_{2}}\rangle$.
This scenario can be described as the action of the $CNOT$ gate on
$|\phi_{G_{1}}\rangle\otimes|\phi_{G_{2}}\rangle$. It is certainly
possible to construct more operations but in principle the $CNOT$
and the 1-quibit operations are enough to represent quantum circuits
\cite{diVin94,diVin95a,diVin96}.

This description showed that our model of a neural network has much
in common with quantum circuits. We don't have in mind that a neural
network and a quantum computer are the same. Here we want to discuss
only the similarities between both descriptions by using our model.

\section{Conclusion}

In this paper we described a qualitative model for neural networks
like the human brain.  Main part of this paper is a new model to
understand the behavior of neural networks. The main idea is a subdivision
of the network into areas with a state given by the signal through
this area consisting of a sum of a ground state $|0\rangle$ and excited
state $|1\rangle$. The interaction between neurons was described
by loops (feedback loops) i.e. by an signal exchange between the neurons.
The state space of all signals was analyzed to get an exceptional
behavior for signals with a large amplitude. These signals forming
a tree which can be used to understand the process of learning in
a neural networks. Furthermore, the signals at the endpoints of this
neural network are related to fractal structures (quasi-Fuchsian groups).

By analyzing the state space in detail, we are able to get a formal
relation to quantum circuits and quantum computing. This relation
is purely formal and has nothing to do with features like entangled
neurons etc. The mathematical description of our model and of quantum
circuits have some aspects in common which was described in the previous
section.

From the philosophical point of view, this approach has global features
which controls the behavior of sub-components. So, the interaction
between neurons given by the feedback loops controls the behavior
of the neurons in the loop. In philosophy, one calls this behavior
top-down causation:\\
\emph{Top-down causation refers to the effects on components of organized
systems that cannot be fully analyzed in terms of component-level
behavior but instead requires reference to the higher-level system
itself.}\\
This model serves as a simple example of this principle. But it showed
that a qualitative model can be more successful then a large scale
simulation. 

%\bibliographystyle{unsrt}
%\bibliography{knots,foliation-gerbes,quantum,diffbib,quantum-theory,exbib_new}

\end{document}